\documentclass[letterpaper,12pt]{article}
\usepackage{graphicx}
\usepackage[latin1]{inputenc}
\usepackage{amsmath}
\usepackage{amsfonts}
\usepackage{amssymb}
\usepackage{amsthm}
\usepackage{bm}
\usepackage{mathrsfs} 
\usepackage[letterpaper,margin=1.5in,top=1.0in,bottom=1.0in]{geometry} 
\usepackage{setspace}
\begin{document}

\section{Introduction}

More than two thirds of the elements in the periodic table have quadrupolar nuclei. Quadrupolar interactions can be as large as or larger than chemical shift and dipolar interactions. Thus, quadrupolar nuclei typically exhibit broad spectra which lack resolution. To resolve quadrupolar spectra, there exist a number of methods, most notably multiple quantum NMR experiments. Multiple quantum experiments are so named because they take advantage of the multiple spin states that characterize quadrupolar nuclei and the coherences that can be created between spin states. The first multiple quantum experiment was performed in 1995 \cite{Fry}. This experiment was called multiple quantum magic angle spinning (MQMAS) and utilized symmetric spin state transitions. Another multiple quantum experiment taking advantage of asymmetric spin state transitions, termed the satellite transition magic angle spinning (STMAS) experiment was introduced in 2000 \cite{Gan}. Although STMAS provides enhanced sensitivity over MQMAS, it has not attracted the same level of interest \cite{Ash09}.

STMAS experiments have an advantage over MQMAS of generally having greater efficiency \cite{Ash04}. Hajjar et al. have studied the effects of phase cycling in MQMAS experiments \cite{Haj}. Though satellite transition experiments have been applied to a variety of spin systems with very good increase in resolution, there have not been any studies in phase cycling effects on the spectrum in these experiments. In this work, I study effects of phase cycling in a satellite transition experiment on a $^{17}$O enriched sample of silica.


\section{Theory}

A perturbation expansion of the quadrupolar Hamiltonian given by Trebosc, et al. \cite{Treb}, written in terms of frequency differences of energy levels is useful for pointing out important features of multiple quantum experiments on quadrupolar nuclei
\begin{align}\label{Eq:QuadHamPert}
\nu^{(1)}_{m,n} &= \frac{\nu_Q(m^2-n^2)}{4}(3 \cos{\beta_R}^2 - 1)d^2_{00}(\chi) \\
\nu^{(2)}_{m,n} &= \frac{\nu_Q^2}{5040\nu_0}(-168[m(S(S+1)-3m^2)-n(S(S+1)-3n^2)] \notag \\ & -60[m(8S(S+1)-12m^2-3)-n(8S(S+1)-12n^2-3)]d^2_{00}(\chi)d^2_{00}(\beta_R) \notag \\ & + 36[m(18S(S+1)-34m^2-5) - n(18S(S+1)-34n^2-5)]d^4_{00}(\chi)d^4_{00}(\beta_R) \label{Eq:QuadHamPert2}
\end{align}
where $m$ and $n$ are the z-components for the states of total nuclear spin $S$, $\beta_R$ is the angle relating the quadrupolar principle axis system (PAS) to the sample rotation frame, $\chi$ is the angle between the static $\bm{B}$ and the sample rotation frame, $\nu_Q$
 is the quadrupole characteristic frequency, and $\nu_0$ is the Larmor frequency of the nucleus. The $d^l_{j \, k}$ are the reduced Wigner rotation matrices. Eqs. \ref{Eq:QuadHamPert} and \ref{Eq:QuadHamPert2} hold when the asymmetry parameter $\eta$ is zero. Taking $\eta=0$ applies to situations where the electric field gradient (EFG) is symmetric about the transverse PAS coordinates. Symmetric crystal structures give $\eta$ values very close to zero \cite{Slich}. In this work, I look at SiO$_2$ crystals with tetrahedral symmetry which makes this assumption valid.

The expansion above shows a couple of important spectral characteristics. First, for rotation about the magic angle, $d^2_{00}(\chi) = (3\cos{\theta_m}^2 -1)/2 = 0$. These terms drop out for sample spinning. Second, for $|m|=|n|$ the first order frequency $\nu^{(1)}_{m,n}$ drops out. For spinning at an appropriate angle, $d^4_{00}(\chi) = (35\cos{\chi}^4 -30\cos{\chi}^2 + 3)/8 = 0$. However no rotation about a single angle will give $d^4_{00}(\chi) = d^2_{00}(\chi) = 0$. Reduction of the $d^4_{00}(\chi)$ term in the second order frequency $\nu^{(2)}_{m,n}$ is the objective of performing quantum coherence experiments.

\begin{figure}[!h]
 \centering
 \scalebox{.5}{\includegraphics{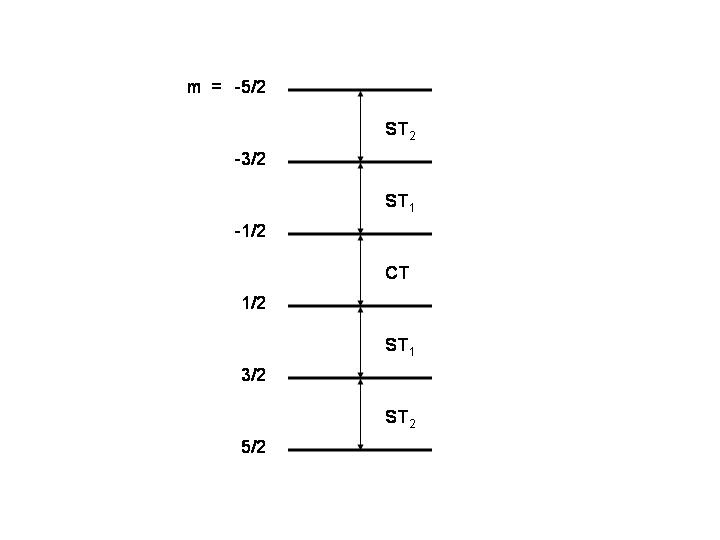}}\\
 \caption{Energy level diagram for a spin $\frac{5}{2}$ nucleus. CT is the central transition. $\text{ST}_{\text{n}}$ in the $\text{n}^{\text{th}}$ satellite transition.}
 \label{fig:st_energy}
\end{figure}

Depending on the spin system, multiple satellite transitions are available. Figure \ref{fig:st_energy} gives an energy level diagram for a spin $\frac{5}{2}$ nucleus with the various possible single quantum satellite transitions. Populations of the states in a system of spins follows a Boltzmann distribution.
Lowest energy states will have the highest populations and thus the largest magnetization. STMAS experiments take advantage of this, transferring magnetization through coherence between the satellite transitions. Ideally, the largest and smallest population magnetizations are transferred to the central transition (CT).


\section{Methods}

\begin{figure}[!h]
 \centering
 \scalebox{.5}{\includegraphics{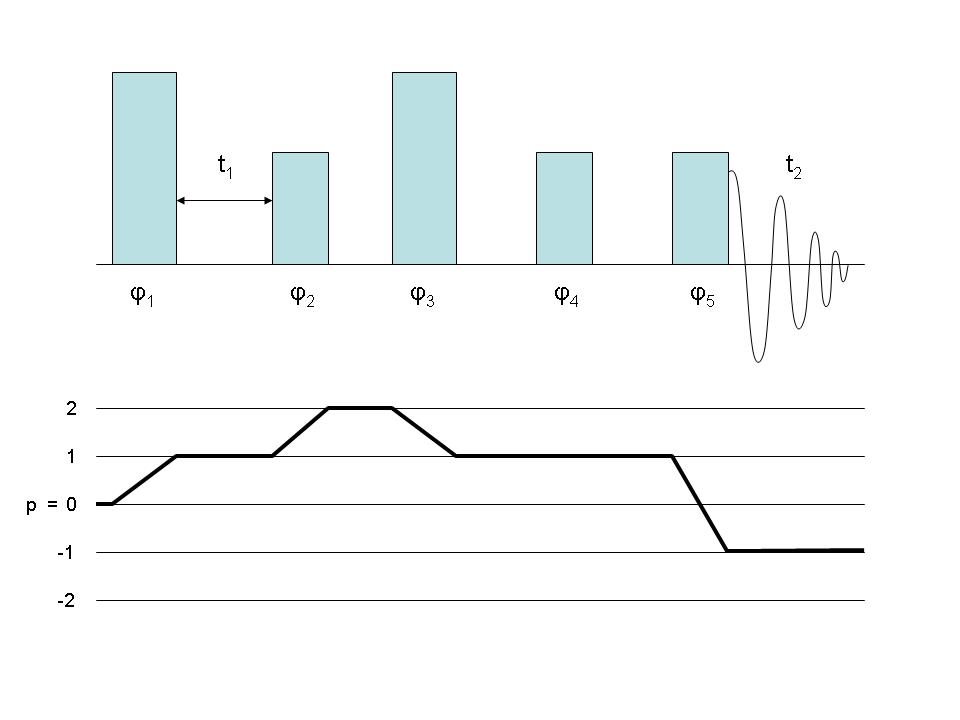}}\\
 \caption{The 1QSTMAS pulse sequence and coherence transfer pathway. The $\phi_i$ are the pulse phases about the z-axis. The $t_1$ variable is incremented to give a 2-dimensional spectrum.}
 \label{fig:1QSTMAS_pulse}
\end{figure}

The pulse sequence and coherence transfer pathway for a single quantum STMAS (1QSTMAS) experiment is shown in Figure \ref{fig:1QSTMAS_pulse}.
The phases $\phi_i$ are determined using the Equations \ref{Eq:COrot} and \ref{Eq:COselect}.
\begin{align}
O(\phi) \rho^p_0 O^{-1}(\phi) &= e^{-i \phi \hat{I}_z} O e^{i \phi \hat{I}_z} \rho^p_0 e^{-i \phi \hat{I}_z} O^{-1} e^{i \phi \hat{I}_z} \notag \\ 
                            &= e^{-i \phi \hat{I}_z} O \rho^p_0 e^{i p \phi} O^{-1} e^{i \phi \hat{I}_z} \notag \\
                            &= e^{-i \phi \hat{I}_z} \sum_n \rho^{p_n} e^{i \phi \hat{I}_z} e^{i p_0 \phi} \notag \\
                            &= \sum_n \rho^{p_n} e^{-i (p_n - p_0) \phi} \label{Eq:COrot}
\end{align}
\begin{align}\label{Eq:COselect}
\Delta p_i^{(selected)} &= \Delta p_i^{(desired)} \pm nN_i, \qquad n = 0, 1, 2,...
\end{align}
To ensure the desired coherence order pathway during observation, the phase of the receiver is also cycled according to \cite{Ernst}
\begin{align}\label{Eq:rec_phase}
\phi_R &= -\sum \Delta p_i \phi_i
\end{align}
For a 1QSTMAS pulse sequence with 4 phases for the first, second, and fifth pulses keeping the third and fourth pulses at a constant phase, to attain one complete signal, the sequence must be performed $4^3$ times.

The observed signal $s(t_1,t_2)$ is 2-dimensional
\begin{align}
s(t_1,t_2) & \propto e^{-i\nu^{(2)}_{\pm \frac{3}{2},\pm \frac{1}{2}}t_1}e^{-i\nu^{(2)}_{- \frac{1}{2},+ \frac{1}{2}}t_2}
\end{align}
The form of the second order term in Equation \ref{Eq:QuadHamPert2}
involves two terms, assuming magic angle spinning. The first term is a constant. The second term gives the fourth order broadening. It is the second term that multiple quantum experiments seek to remove. As seen in the 2-dimensional signal, appropriate choices of $t_2$ given a $t_1$ will result in only a constant exponential term. This ``refocusing'' of the signal to remove second order broadening is the means by which multiple quantum experiments attain spectra with improved resolution in the second dimension.

\begin{figure}[!ht]
 \centering
 \scalebox{.5}{\includegraphics{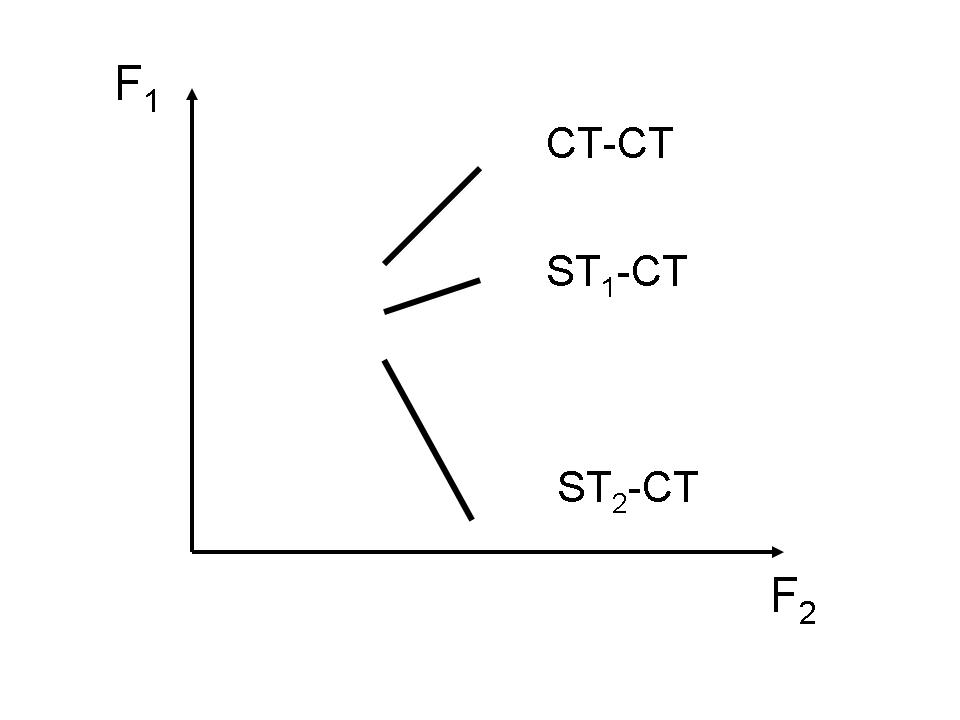}}\\
 \caption{Schematic of possible results from a 1QSTMAS experiment performed on a spin $\frac{5}{2}$ nucleus. F$_1$ and F$_2$ are the Fourier transformed $t_1$ and $t_2$ dimensions. The slope of lines in the spectrum are calculated as the ratio, $R$, of second order rank 4 broadenings of the transitions involved. Thus $R=1$ for CT-CT, $R=7/24$ for ST$_1$-CT, and $R=-11/6$ for ST$_2$-CT.}
 \label{fig:STMAS_schematic_result}
\end{figure}

\begin{figure}[!ht]
 \centering
 \scalebox{.5}{\includegraphics{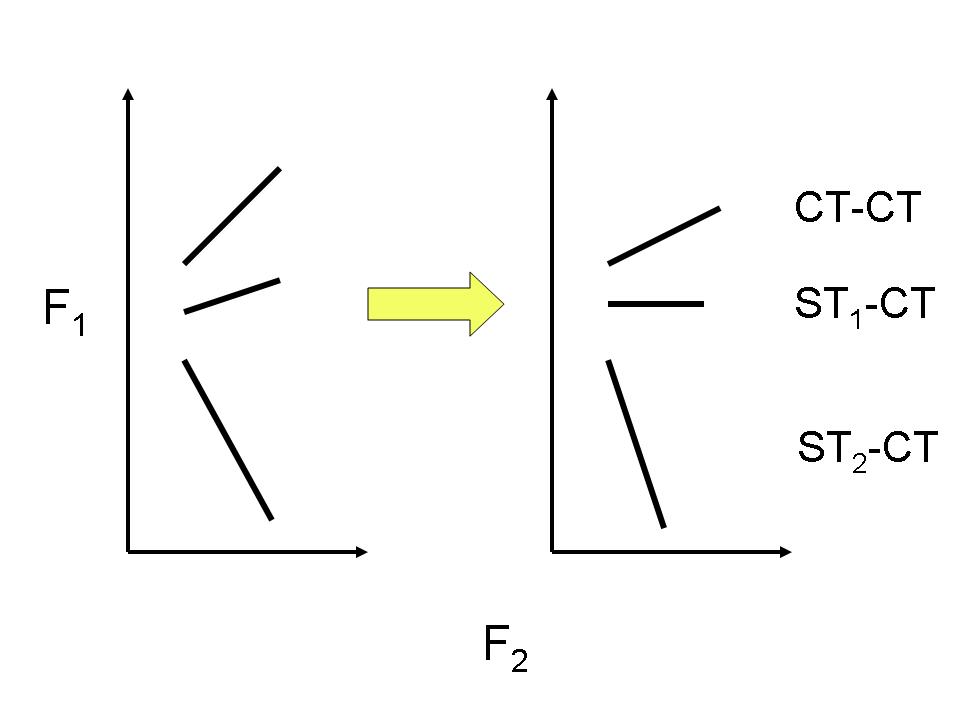}}\\
 \caption{Shearing of STMAS spectrum. The spectrum has been rotated so that the ST$-1$-CT portion gives the isotropic projection on the F$_1$ axis.}
 \label{fig:STMAS_shear}
\end{figure}

Figure \ref{fig:STMAS_schematic_result} is a schematic representation of possible results from a 1QSTMAS experiment performed on a spin $\frac{5}{2}$ nucleus. Generally, results include contributions from the various transitions. According to Figure \ref{fig:st_energy}, these contributions come from the CT, ST$_1$, and ST$_2$ transitions. 
The slopes of the lines in the spectrum are calculated as the ratio, $R$, of second order quadrupolar broadenings of the transitions involved. Thus $R=1$ for CT-CT, $R=7/24$ for ST$_1$-CT, and $R=-11/6$ for ST$_2$-CT \cite{Gan01, Ash02}. The spectrum is typically sheared according to the $R$ to give the isotropic spectrum in the $F_1$ dimension \cite{Ash04}. Figure \ref{fig:STMAS_shear} shows the effect of shearing for the ST$_1$-CT portion to give an isotropic projection on the F$_1$ axis.
Results of the 1QSTMAS experiment for $^{17}$O in silica are presented in the next section.


\section{Results}

Phase cycling STMAS experiments were performed on $\text{SiO}_2$ enriched to 10\% $^{17}\text{O}$. The sample was graciously provided by Marek Pruski at Iowa State University. All experiments were performed on a 14.09 T magnet with a 3.2 mm probe and 10 kHz sample spinning. Spectra have been referenced to the frequency of 81.312792 MHz. The theory presented in Equation \ref{Eq:COselect} implies that phase cycling of the individual pulses influences the change in coherence orders available in an experiment. Selection of the coherence transfer pathway is determined by Equation \ref{Eq:rec_phase}. The objective is to test whether phase cycling influences coherence orders in the spectrum. A 1QSTMAS experiment was implemented varying the phase of the first, second, and fifth pulses in the pulse sequence shown in Figure \ref{fig:1QSTMAS_pulse}. In this implementation of the pulse sequence, the third pulse is not phase cycled and the fourth pulse does not give a change in coherence order by Equation \ref{Eq:rec_phase}. All spectra presented have been sheared to give the ST$_1$-CT isotropic along the F$_1$ axis, unless otherwise stated.

\begin{figure}[!ht]
 \centering
 \scalebox{.5}{\includegraphics{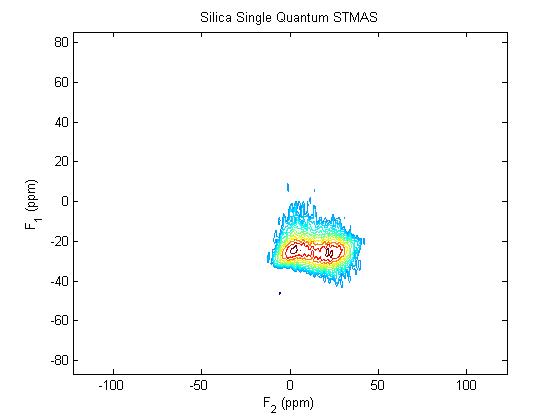}}\\
 \caption{The 1QSTMAS spectra of $^{17}\text{O}$ in 10\% enriched $\text{SiO}_2$. Experiments were performed on a 14.09 T magnet.}
 \label{fig:sio_1QSTMAS_contour}
\end{figure}

\begin{figure}[!ht]
 \centering
 \scalebox{.5}{\includegraphics{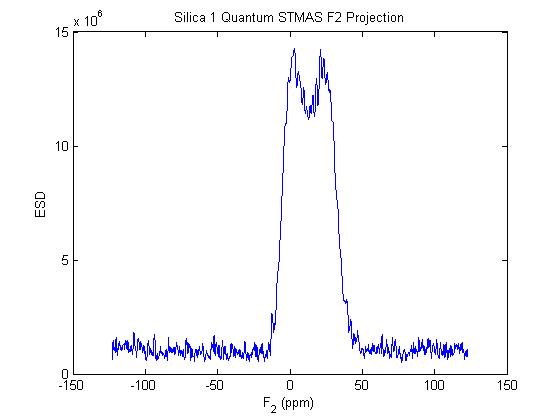}}\\
 \caption{The projection of Figure \ref{fig:sio_1QSTMAS_contour} along the F2 axis.}
 \label{fig:sio_1QSTMAS_F2proj}
\end{figure}

\begin{figure}[!ht]
 \centering
 \scalebox{.5}{\includegraphics{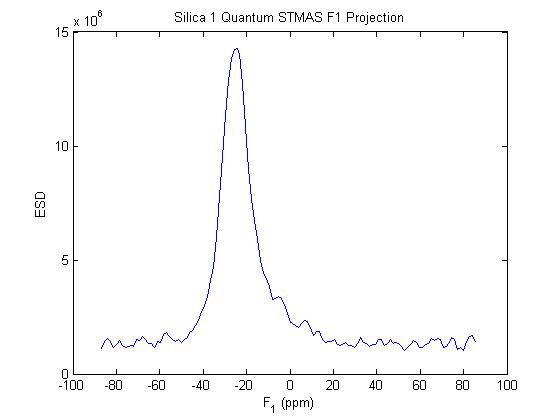}}\\
 \caption{The projection of Figure \ref{fig:sio_1QSTMAS_contour} along the F1 axis.}
 \label{fig:sio_1QSTMAS_F1proj}
\end{figure}

Figure \ref{fig:sio_1QSTMAS_contour} is a contour plot of the result of a 1QSTMAS experiment performed with 4 phases each for the first, second, and fifth pulses. This phase cycling will be considered the standard phase cycling. Figures \ref{fig:sio_1QSTMAS_F2proj} and \ref{fig:sio_1QSTMAS_F1proj} give the projections along the $F_2$ and $F_1$ dimensions. The $F_2$ projection is the spectrum obtained with a reduction in first order broadening from sample spinning. Second order broadening is still present, resulting in what appears to be two peaks in the spectrum. The $F_1$ dimension shows a narrowing of the spectrum from reduction of the second order broadening using the ST$_1$-CT coherences. 

The phase cycle of $\phi_5$ was varied by 1, 2, 4, or 8 phases while phases were cycled normally for all other pulses. The spectral width in $F_2$ is 100 kHz while that in $F_1$ is 10 kHz. For each phase, I performed the experiment twice. Once with the number of acquisitions constant across phase changes (the number of complete phase cycles varied across phase changes), and once with the number of complete phase cycles constant across phase changes (the number of acquisitions varied across phase changes). The metrics used for comparison are the total two-dimensional integrated area of the spectrum and the full width at half max (FWHM) of the $F_1$ projection. The results are summarized in Table \ref{tab:sio_phase_results}.
\begin{table}[!h]
\begin{center}
 \caption{Results of the phase cycling experiment for 1QSTMAS on silica.}
 \begin{tabular}{c c c c c}
  Number of \\ $\phi_5$ Phases & \multicolumn{2}{c}{32 Complete Phase Cycles} & \multicolumn{2}{c}{512 complete acquisitions} \\ \hline
   & Integral ($\times 10^{10}$) & FWHM (ppm) & Integral ($\times 10^{10}$) & FWHM (ppm)\\ \hline
  8 & 11.9 $\pm$ 0.00000170 & 14.24 $\pm$ 0.018 & 1.47 $\pm$ 0.00000447 & 14.92 $\pm$ 0.025\\
  4 & 5.85 $\pm$ 0.00000218 & 12.89 $\pm$ 0.021 & 1.52 $\pm$ 0.00000417 & 13.49 $\pm$ 0.023\\
  2 & 3.01 $\pm$ 0.00000310 & 12.89 $\pm$ 0.020 & 1.41 $\pm$ 0.00000433 & 14.92 $\pm$ 0.021\\
  1 & 1.53 $\pm$ 0.00000431 & 13.50 $\pm$ 0.024 & 1.52 $\pm$ 0.00000421 & 13.49 $\pm$ 0.023 \\
 \end{tabular}
 \label{tab:sio_phase_results}
\end{center}
\end{table}

The integrals in the experiments with 32 complete phase cycles show a trend consistent with the increase in the number of acquistions defined by the phases. For example, the number of acquisitons for a complete phase cycle with $\phi_5 = 1$ is 16 while for $\phi_5 = 2$ is is 32. The number of acquistions for a complete phase cycle is equal to $N_1 N_2 N_3$ where $N_i$ is the number of phases for the $i$-th pulse. Thus the integral values double as the number of $\phi_5$ phases double. Variation in the FWHM over acquisitions are repeatable, but the reason is not clear. These results gerenally correspond to those reported in phase cycling in MQMAS experiments \cite{Haj}. In particular, in STMAS experiments each acquisition adds constructively regardless of phase cycling.

\begin{figure}[!ht]
 \centering
 \scalebox{.5}{\includegraphics{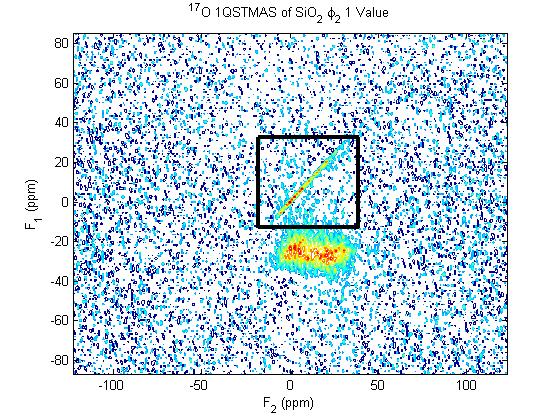}}\\
 \caption{The spectrum for one value of $\phi_2$. The boxed area shows a CT-CT line not apparent in the 4 phase cycle spectrum.}
 \label{fig:sio_414}
\end{figure}

\begin{figure}[!ht]
 \centering
 \scalebox{.5}{\includegraphics{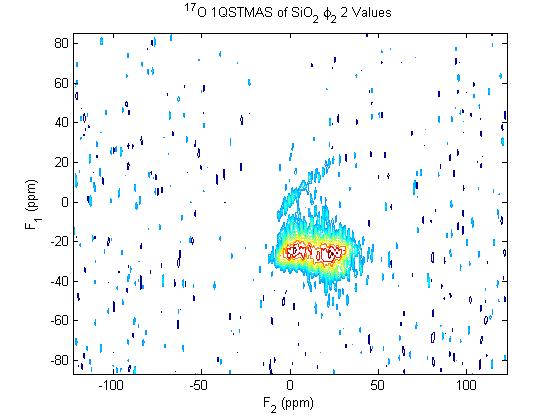}}\\
 \caption{The spectrum for two values of $\phi_2$. The CT-CT coherence line is still apparent.}
 \label{fig:sio_424}
\end{figure}

Phase cycling on $\phi_1$ and $\phi_2$ produced altogether different results. The phase cycle for each was varied by 1 and 2 phases while all other phases were cycled normally. Data for each experiment consist of 32 complete phase cycles. Results for 1 value of $\phi_2$ are shown in Figure \ref{fig:sio_414}. The spectrum shows the appearance of a CT-CT coherence line which does not show up in the 4 phase cycle spectrum. Likewise, Figure \ref{fig:sio_424} shows a CT-CT line, though with less magnitude.

\begin{figure}[!h]
 \centering
 \scalebox{.5}{\includegraphics{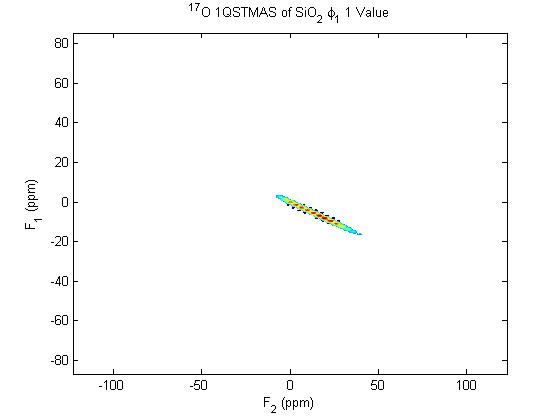}}\\
 \caption{The spectrum for one value of $\phi_1$.}
 \label{fig:sio_144}
\end{figure}

\begin{figure}[!h]
 \centering
 \scalebox{.5}{\includegraphics{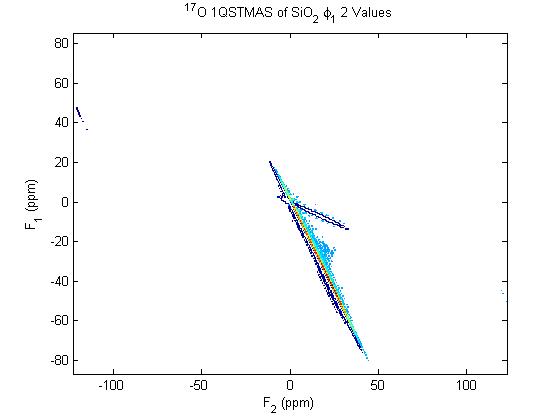}}\\
 \caption{The spectrum for two values of $\phi_1$.}
 \label{fig:sio_244}
\end{figure}

Figures \ref{fig:sio_144} and \ref{fig:sio_244} show results for $\phi_1$ phase cycling of 1 and 2 values, respectively. These spectra show lines that do not appear to correspond to CT-CT, ST$_1$-CT, or ST$_2$-CT coherences. Furthermore, the spectra appear to be shifted in frequency along the F$_1$ axis compared to the 4 phase cycle spectrum of Figure \ref{fig:sio_1QSTMAS_F2proj}.

\section{Conclusion}

Results for phase cycling of the first and second pulses in the 1QSTMAS experiment contrast with results obtained by phase cycling in MQMAS experiments performed by Hajjar, et al. \cite{Haj}. Experiments performed on the 1QSTMAS experiment in this work show that phase cycling may or may not affect the spectrum. In particular, a decrease in FWFM of up to 14\% can be achieved. Furthermore, cycling $\phi_2$ may result in the CT-CT coherence appearing in the spectrum as well as the ST$_1$-CT coherence. Cycling of $\phi_1$ results in as yet undetermined lines appearing in the spectrum. Although it has been stated that CT-CT transitions cannot be removed from the spectrum \cite{Ash04}, clearly this study shows that phase cycling does have an influence over appearance of these transitions. A possible explanation might be that phase cycles of 1 and 2 produce a wider variety of coherence orders than the 4 phase cycle. According to Equation \ref{Eq:COselect}, a phase cycle of 1 allows $0, \pm 1, \pm 2,$... coherence orders. A phase cycle of 2 allows ...-3,-1,1,3,... coherence orders. On the other hand, a 4 phase cycle would only allow ...-4,1,5,... coherence orders. Receiver phase cycling defined in Equation \ref{Eq:rec_phase} my not be able to completely remove the allowed coherence orders for phase cycles of 1 and 2.

\begin{singlespace}
\bibliographystyle{plain}
\bibliography{phase_cycling_in_stmas_bib}
\end{singlespace}

\end{document}